\documentstyle[11pt,aaspp]{article}

\begin{document}
\title{Testing the Accuracy of Redshift Space Group Finding Algorithms}
\author{James J. Frederic}
\affil{Department of Physics, MIT 6--110, Cambridge, MA 02139}

\begin{abstract}
Using simulated redshift surveys generated from a high resolution N-body
cosmological structure simulation, we study algorithms used to identify groups
of galaxies in redshift space.  Two algorithms are investigated; both are
friends-of-friends schemes with variable linking lengths in the radial and
transverse dimensions.  The chief difference between the algorithms is in the
redshift linking length.  The algorithm proposed by Huchra \& Geller (1982)
uses a generous linking length designed to find ``fingers of god'' while
that of Nolthenius \& White (1987) uses a smaller linking length to minimize
contamination by projection.

We find that neither of the algorithms studied is intrinsically superior to the
other; rather, the ideal algorithm as well as the ideal algorithm parameters
depend on the purpose for which groups are to be studied.  The Huchra/Geller
algorithm misses few real groups, at the cost of including some spurious groups
and members, while the Nolthenius/White algorithm misses high velocity
dispersion groups and members but is less likely to include interlopers in
its group assignments.

In a companion paper we investigate the accuracy of virial mass estimates
and clustering properties of groups identified using these algorithms.
\end{abstract}


\section{Introduction}
Single bright galaxies are numerous and trivially easy to identify,
and are hence well studied.  Clusters of hundreds or even thousands
of galaxies, while far less abundant than individual galaxies, are
still relatively easy to identify
due to their high projected density of galaxies.  They are also well
studied.  Between these two scales of structure are groups, which,
though difficult to define, can be thought of as those dynamical associations
of galaxies on scales smaller than those of clusters.  A useful working
definition
of a group is an enhancement of either number, luminosity or mass
density above a certain threshold.  This definition includes clusters,
treating groups and clusters as similar phenomenona which
differ quantitatively, not qualitatively.  It is a natural definition
when clustering is hierarchical.  In this work we adopt the definition
used by Huchra \& Geller (1982, hereafter HG82) and several subsequent workers,
that
groups are number density enhancements in redshift space.

The initial difficulty in performing any study of galaxy groups is their
accurate
identification.  A few errors in assigning membership to a
cluster of over 100 galaxies will presumably have only slight effects on
a dynamical analysis of the cluster.  Typical groups have only a
few members, so great care must be taken in determining group membership.
Toward these ends, astronomers have developed various techniques for
identifying groups of galaxies.  Early efforts either produced group catalogs
based on subjective definitions and with ill defined sampling
(de Vaucouleurs 1975) or depended
solely on two dimensional positional data (Turner \& Gott 1976).  With
the availability of large, complete redshift surveys, it became possible
to generate group catalogs with objective and well understood selection
criteria (Press \& Davis 1982; HG82;
Geller \& Huchra 1983, hereafter GH83;
Nolthenius \& White 1987, hereafter NW87;
Ramella, Geller \& Huchra 1989, hereafter RGH89).
Studies of the groups in these catalogs are limited by the difficulty in
assessing the uncertainties in group identification.

In this paper we are concerned with identifying galaxy groups and
determining the accuracy of the identification techniques.
A second paper (Frederic 1994, hereafter Paper II) focuses on
the internal properties of groups (masses, mass to light ratios,
internal velocity dispersions, etc.) as well as their clustering properties.
We study two algorithms for identifying
galaxy groups from redshift surveys.  The first was introduced
by HG82 and used by the same
authors in GH83 and by RGH89.  The second was
proposed by NW87 as
an improvement to the first.
These algorithms or a hybrid of the two have been used to construct
real group catalogs (HG82; GH83; NW87; RGH89; Nolthenius 1993) and
have recently been applied to simulated redshift surveys to discriminate
between cosmological models (Nolthenius, Klypin, \& Primack 1994).
We apply these algorithms to
simulated redshift survey data obtained from an N-body
simulation performed by Gelb (1992) in order to test the accuracy
of the group identification.

Although the group finding algorithms we study have been tested in the past
by application to simulated data, the results of those tests carry with
them all the uncertainties about the accuracy of the simulations themselves.
By using an improved simulation for our tests we hope to significantly
strengthen our confidence in the accuracy of our group studies.
When NW87 and Press \& Davis (1982) each applied their
techniques to the results
of N-body experiments in order to test the overall accuracy of their
group finders, they were limited by the state of art in N-body experiments
to particle masses comparable to or even greater than the mass of a single
galaxy.  Moore, Frenk \& White (1993) suffered from the same limitation
when they looked for groups
in another N-body simulation in their study of the group luminosity
function.  The simulation used here improves over these
in mass and force resolution, and the additional dynamical information
it provides allows improved modeling of the luminosities of the simulated
galaxies.

The primary advantage of our improved resolution is in
the identification of simulated ``galaxies.''  Unlike those studies mentioned
above,
we identify galaxies from the evolved density field.  And because our galaxies
have a range of masses, the masses of our groups are not coarsely
quantized.  We thereby decouple the distributions of group mass
and group richness (number of member galaxies).  By locating galaxies
from the evolved simulation, we also have more confidence in our derived
spatial distribution of galaxies and groups.
Although this simulation assumes a particular cosmological
model (cold dark matter, or CDM), it will be used here primarily to test
deductions
about groups
based on redshift information rather than to study group properties in a CDM
universe.

Because the properties of simulated groups are expected to be sensitive
to the manner in which galaxies are identified in the simulation,
we use an improved algorithm known as DENMAX
for identifying galaxies as density enhancements in the
mass distribution of the
simulation.  This improved galaxy identification procedure
compares favorably to the common friends-of-friends method
(Davis et al. 1985; Brainerd \& Villumsen 1992)
and to the peak particle method, which involves tagging
particles near peaks in the initial density field as ``galaxy particles''
and treating those and only those particles as galaxies throughout
the simulation (Katz, Quinn, \& Gelb 1993; Gelb \& Bertschionger 1994b).

We attempt to construct simulated redshift catalogs with clustering
properties similar to those of the real universe.  We compare our catalogs
to
the first $6^{\circ}$ declination slice of the Center for Astrophysics (CfA)
redshift survey extension complete to $m_B=15.5$ (Huchra {et al.} 1990).
The data were obtained electronically through
the Astronomical Data Center of the National Space Science Data
Center/World Data Center A for Rockets and Satellites at NASA Goddard Space
Flight Center.  We used the February 1992 version of the catalog,
ADC catalog number 7144 (Huchra {et al.} 1992).

Section 2 describes the N-body experiment and DENMAX, the method
of galaxy identification in the simulation, as well as the method
by which magnitude limited redshift catalogs were generated from
the simulation.  Specific group finding algorithms are discussed
and their accuracy studied in section 3.
The final section presents conclusions.

\section{Constructing the Simulated Redshift Survey Catalogs}
Constructing simulated redshift space galaxy catalogs
requires an N-body simulation and a method for identifying
``galaxies'' from the N-body data.  (I will use the term
``halo'' from now on to refer to clumps of dark matter in
the simulation, and reserve ``galaxy'' for the real data.)
These procedures should produce
catalogs which are as statistically similar to real data as possible,
particularly in regards to clustering on scales important for groups.

\subsection{The Simulation}
We use a simulation performed by Gelb (1992)
for his doctoral thesis.  This simulation uses a
modified version of Couchman's (1991) particle-particle--particle-mesh (P$^3$M)
algorithm
to dynamically evolve a gas of collisionless
dark matter particles with standard CDM initial
conditions in an expanding universe with a present day Hubble
parameter $H_0=100h$ km s$^{-1}$ Mpc$^{-1}$, with $h=0.5$.
The volume simulated is a periodic cube with sides of
length 5000 km s$^{-1}$ ($50h^{-1}$ comoving Mpc) which
was evolved for 1200 timesteps to a linear amplitude
$\sigma_8=1$, the ``unbiased'' amplitude consistent with the COBE
quadrupole measurement of the microwave background radiation
anisotropy.
Because the simulated volume is periodic, we are free to
enlarge it by stacking replicas of the fundamental volume against each other.
In this way we are able to construct simulated redshift survey catalogs
to a depth of 15000 km s$^{-1}$ from a 5000 km s$^{-1}$ simulation.
In this work
we analyze the simulation output corresponding to $\sigma_8=0.7$,
an epoch which the analysis by Gelb (1992) shows to be in better
agreement with the observations with respect to the numbers
of massive halos and the halo velocities.  As we are
exploring redshift space, it is is especially important
that our simulated galaxy catalogs have velocities which
match the data reasonably well in a statistical sense.
The simulation evolves 144$^3$ ($\approx 3$ million)
particles, each weighing $1.16\times 10^{10}h^{-1}M_{\odot}$.  A Plummer law
with $32.5h^{-1}$ kpc (comoving) softening radius was used for the force
calculation.

Previous simulations, specifically those used by NW87 and Moore {et al.} (1993)
to test group finding algorithms, used more massive particles
($3.4 \times 10^{12}h^{-1}M_{\odot}$ to $1.9 \times 10^{13}h^{-1}M_{\odot}$ and
$6.2 \times 10^{12}h^{-1}M_{\odot}$, respectively) and so were
forced to identify one or more ``galaxies'' with single particles.
NW87 used softening radii
ranging from $350$ to $660h^{-1}$ comoving kpc, and Moore {et al.} (1993)
used a softening length of $562.5h^{-1}$ comoving kpc.  This
severely limits the degree of confidence which can be placed in their
group analysis, since NW87 found in their analysis of the first
CfA survey that real groups have
typical radii of over $600h^{-1}$ kpc.
(Precisely defined measures of group size are given in Paper II and NW87.)
Since forces are not being accurately calculated between pairs
of particles separated by less than about two softening lengths, group dynamics
at these scales cannot be accurately modeled.

\subsection{Clustering statistics}
Before constructing our catalogs we develop statistics for comparing
them to the CfA data.  Because clustering on the scales of groups
is an important feature to mimic, we use the two point correlation
function in redshift space, estimated as
\begin{equation}
\xi(s)={n_R \over n_D}{N_{DD}(s) \over N_{DR}(s)} -1 \,,
\label{xis}
\end{equation}
\begin{equation}
s={(V_i^2 + V_j^2 -2 V_i V_j \cos \theta_{ij})^{1/2} \over H_0} \,,
\end{equation}
where $V_i$ and $V_j$ are the radial velocities of two galaxies with angular
separation $\theta_{ij}$ and $H_0$ is Hubble's constant.
In order to account for edge effects, a catalog of randomly distributed
points with geometry and selection function identical
to the real or simulated data must
be generated.  Then $n_D$ and $n_R$ are the number of points in the data
and the random catalog, respectively, $N_{DD}(s)$ is the number of pairs
in the data separated by redshift distance $s$, and $N_{DR}(s)$ is the number
of pairs, one from the data and one from the random catalog, separated
by $s$.  Our random catalogs used in calculating the galaxian correlation
function contain approximately the same density of points as do the
simulated data.
In order to minimize
the statistical noise from the random catalog, we compute $\xi(s)$ for
ten separate simulated catalogs corresponding to ten different observers,
using a different random catalog for each observer.  These results are then
averaged.

We separate the contributions to $\xi(s)$ from the
radial separation $\pi=|V_i - V_j|$ and the transverse separation
$H_0 r_p= (V_i + V_j) \tan (\theta_{ij} / 2)$ by using $\xi(r_p,\pi)$,
which is determined analogously to $\xi(s)$ by replacing the single
parameter $s$ with the parameters $r_p$ and $\pi$ in equation (\ref{xis}).
Contours of constant $\xi(r_p,\pi)$ show extension in the velocity
direction when the groups are distorted into the so-called ``fingers
of god'' along the line of sight in redshift space maps.

\subsection{Identifying simulated ``halos''}
When our evolved simulation is viewed in projection, the eye immediately
picks out clumps of particles
as distinct objects.  We have developed an algorithm known as DENMAX
which mimics the choices one makes with one's eye.  The density field
is evaluated on a very fine grid using TSC interpolation
(Hockney \& Eastwood 1981) from the particle
positions.  It is then convolved with a gaussian filter with a
radius of three grid units, so as to eliminate grid effects.
For our purposes, a smoothing length of one thousandth of the edge
of the simulation volume, corresponding to $50h^{-1}$ kpc and
appropriate for identifying galactic sized density enhancements,
was used.  Smoothing this small, with three grid points per smoothing
radius, means the density field is being calculated at $3000^3$ grid
points.  In order to stay within the memory constraints of our
computers, we perform the density calculation on multiple subvolumes
of the total simulation.
In high density subvolumes, the convolution is performed
in Fourier space to take advantage of the speed of the Fast Fourier
Transform.  In low density subvolumes, where the unsmoothed density
field is zero over much of the volume, the FFT is inefficient, and
smoothing is performed by convolution in real space.  Due to the use
of such a fine grid, the density calculation is by far the most
computationally expensive portion of DENMAX, requiring about 60 hours
to analyze a $144^3$ particle simulation on a Convex C3880.

Once the smoothed density field
has been calculated, particles are moved up the local gradient of this
field into the density peaks by integrating the
equation $d \vec x / dt = \vec \nabla \rho / \rho$.
Here $t$ is a fictitious time variable, and $dt$ is
just a coefficient for the calculated displacement of a particle.
To calculate the optimal step
size we use the following logic.  The gradient in the smoothed density
field is steepest near an isolated mass concentration.  Here the shape
of the density field approaches pure gaussian, and knowledge of the height
and slope at a point is equivalent to knowing the location of the
peak.
\begin{equation}
\rho(\vec x)={m \over (2\pi\sigma^2)^{3/2}} \exp\left[-{(\vec x - \vec x_0)^2
\over 2\sigma^2}\right] \,,
\end{equation}
\begin{equation}
{\vec \nabla \rho \over \rho}(\vec x) = -{\vec x - \vec x_0 \over \sigma^2}
\end{equation}
So in this most extreme case, the optimal step size is $|{\vec \nabla \rho
\over \rho}| \sigma^2$, which would take the particles directly to the
peak in one step.  To handle the general case efficiently, we multiply this
step size by a scale factor of order unity, choosing this factor
to maximize the speed of convergence of the algorithm without causing
particles to overshoot and oscillate across the peaks.  In addition,
the step distance
is limited to be no more than one fourth of one grid spacing, to make sure
that the steps are much smaller than the scales over which the field changes
appreciably and to ensure that particles do not wander into the
wrong density peak.  Our procedure is equivalent to dragging
particles through a highly viscous fluid toward the density peaks,
with individual timesteps guaranteeing small
physical steps for each particle.
This portion of the DENMAX algorithm vectorizes well
and requires only a small fraction of the total run time.
Once the algorithm converges, the particles at each peak can be easily
grouped by a friends-of-friends type algorithm with a very small linking
length.

Unlike the peak particle method, DENMAX makes no
{\it a priori} assumptions about the sites of halo formation.
This is important because initial peak particles do not
necessarily make good choices for halos.
Katz {et al.} (1993) show that the final positions of
particles near peaks in the initial density field do not correspond
well to the positions of peaks in the evolved density field.
DENMAX, on the other hand, identifies halos with peaks
in the evolved density field.  Friends-of-friends methods also apply to the
evolved density field, but they sometimes cause distinct density peaks
to be linked together.  Gelb \& Bertschinger (1994a) study differences
between an earlier version of DENMAX and friends-of-friends.
Other methods presented by Klypin et al. (1993) identify halos
with peaks in the evolved density field defined on a regular grid, but use
somewhat {\it ad hoc} methods for defining the mass of each halo.
Although not as simple in its implementation as other methods of
halo identification, DENMAX is conceptually straightforward,
having the density smoothing length as its only free parameter,
and leads to reasonable assignments of particles to halos as well as
a one to one correspondence between halos and peaks in the evolved density
field.

\subsection{Illuminating the simulated halos}
Once halos have been located by DENMAX, they must be assigned luminosities
so that an apparent magnitude limit may be applied to our simulated catalogs.
Gelb (1992) illuminated halos by calculating their circular
velocities at a fixed radius or their one dimensional internal velocity
dispersions, and then assuming a Tully-Fisher type relationship for
70\% of the halos and a Faber-Jackson relationship for the other 30\%.
He found that the resulting luminosity function matched observations
in the intermediate luminosity range, but differed significantly
at both the bright and faint ends.  A more recent simulation
by Katz, Hernquist, \& Weinberg (1992) which includes
gas dynamics in a CDM universe obtains a more accurate match to the
observed luminosity function over the scales they probe, which again
are of intermediate range.   They claim to see evidence that
better resolution would lead to a problem with too many
faint ``galaxies,'' but that the inclusion of still more
detailed gas and radiation physics may solve this problem.
Efstathiou (1992) suggests
that the faint end slope of the luminosity function is extremely
sensitive to any ionizing background radiation.

Because we are concerned here with galaxy groups and clustering, and not
with any specific cosmological model, we will force our luminosity
function to match the observations, which have been fit to the Schechter
(1976) form
with the following parameters by de Lapparent, Geller, \& Huchra (1988):
\begin{eqnarray}
M_{B(0)}^*=-19.15+5\log h \,,
\nonumber
\end{eqnarray}
\begin{equation}
\phi^*=0.025h^{3}{\rm galaxies ~Mpc}^{-3} \,, \qquad \alpha=-1.2 \,.
\label{schech}
\end{equation}

NW87 and Moore {et al.} (1993) forced their luminosity function
to match observations by randomly sampling their
desired luminosity function for each ``galaxy.''
Because our halos are distinguishable, we have more information
to use in assigning luminosities.  We have chosen to select luminosities
from the observed luminosity function while
preserving the rank order of halo circular velocities,
\begin{equation}
V_{\rm circ}=\sqrt{GM(<R)/R} \,,
\end{equation}
where the mass $M(<R)$ interior to halocentric radius $R$ includes
only the gravitationally bound particles.
That is, the $n$th
brightest halo has the $n$th highest circular velocity.
The procedure whereby gravitationally
unbound particles are removed is described in Gelb \& Bertschinger (1994a).

Due to the $32.5h^{-1}$ kpc force softening used in the simulation,
our halos are not
as condensed as real galaxies.  However, the circular velocities
are flat at large radii.  Gelb (1992)
has shown that most of our halos have flat rotation curves at a radius
of $100h^{-1}$ kpc.  We adopt this value in computing $V_{\rm circ}$ to
construct our ``fewest assumptions'' catalog (no cluster breakup
scheme or linearly added long wavelength power; these will be
described later), which we will call our
raw simulated halo catalog.  We construct both an apparent magnitude limited
and an absolute magnitude limited version of the halo catalog; this will
be described further below.
For the apparent magnitude ($m_B=15.5$) limited version of this catalog and
those introduced
later, Table \ref{galcatalogs} gives the number of halos, their median
redshift,
their three dimensional {rms} peculiar velocities,
and parameters for fitting the correlation function $\xi(s)$ to
the form
\begin{equation}
\xi(s)=(s / s_0)^\gamma
\end{equation}
over the range $1h^{-1} \leq s \leq 10h^{-1}$ Mpc.
For the raw and breakup halo catalogs, quantities quoted are the mean
$\pm$ one standard deviation for our ten simulated catalogs.

We chose to base our luminosities on circular velocity, as opposed
to some other measure of mass or size.  The primary motivation
for this choice is the fact that circular velocity correlates
well with luminosity in spiral galaxies.  Luminosity is correlated
with internal velocity dispersion in elliptical galaxies and, for
simple models, $\sigma$ increases
monotonically with $V_{\rm circ}$ (e.g., for an isothermal sphere, $V_{\rm
circ}=\sqrt{2}\sigma$).
Spirals and ellipticals
maintain separate correlations with
different zero points.
The best solution
to this problem requires that we treat some halos as spiral galaxies
and the rest as ellipticals.  Basing luminosity on $V_{\rm circ}$ alone
is equivalent to treating all of our halos as spiral galaxies.
In order to test whether distinguishing halos by type is important,
we conservatively assume
a high elliptical fraction (70\% spiral, 30\% elliptical).
We then randomly selected a type for each of our halos,
applied the Tully-Fisher or Faber-Jackson relation as appropriate,
and then rescaled the luminosities, preserving their order,
to fit the observed luminosity function.
The changes in $\xi(s)$, halo and group numbers and the calculated virial
mass to light ratios of groups (analyzed in Paper II) which result
from this random type based illumination technique are all small compared
to the uncertainties in the CfA data.  We therefore prefer the simpler
illumination technique based on $V_{\rm circ}$ alone.

Perhaps a more informed choice of halo type could
be made on the basis of the density-morphology relation of Dressler (1980),
which states that the elliptical fraction rises in regions dense with
galaxies.  Because ellipticals tend to be fainter than spirals for a given
$V_{\rm circ}$ (if $V_{\rm circ} = \sqrt{2} \sigma$),
a procedure based on this assumption would make our group
and cluster members fainter and would thereby lower $\xi(s)$, which is
already lower than the observations.  Yet another alternative, using the
cluster-centered radius vs. morphology relation proposed by Whitmore
\& Gilmore (1991) would lower $\xi(s)$ for the same reason, and would
require some kind of (possibly non-unique) iterative scheme to
assign luminosities and identify groups simultaneously.

As an exploration of different methods of assigning
luminosities to our simulated halos,
we generated catalogs with luminosities based either on total halo
mass or on total bound halo mass ($V_{\rm circ}$, remember, is
equivalent to the bound halo mass within $100h^{-1}$ kpc).  The correlation
function
$\xi(s)$ of the bound mass catalog matches that of the $V_{\rm circ}$ catalog,
while the catalog based on total mass shows a slightly higher correlation.
This can be understood in light of the work of Kaiser (1984),
who showed that for a gaussian random field, higher peaks are more
strongly correlated.  By construction, our simulation began from
a gaussian random initial density field.  Although the final density field
was not gaussian random, it was shown by Gelb (1992) to display
stronger correlations between more massive
objects.  Removing gravitationally
unbound particles from the halos or applying a radius cut adds a
dispersion to this relationship, causing some halos to be less brightly
illuminated than they would be otherwise.  These halos may then fall below
the magnitude limit of the catalog, thus decreasing the overall
correlation function.  This fact makes the use of total mass (rather
than $V_{\rm circ}$) attractive, as it would make $\xi(s)$ for
the simulations agree better with the CfA data.
However, much of this total mass is relatively
far from the density peak where the luminous galaxy presumably forms.
Our goal here is to generate a catalog with clustering properties similar to
the
real data, but by a reasonable scheme motivated by observations.  Because
luminosity correlates well with circular velocity in the real universe,
we will use the simple $V_{\rm circ}$-based luminosity
ranking to generate our simulated redshift catalogs.

\subsection{Overmerging}
A major limitation of the simulation used in this work has
been dubbed the overmerging problem.  Colliding groups of particles
rapidly merge, erasing almost all substructure.  The result is that the
simulation
contains too many high mass halos (Gelb 1992, Gelb \& Bertschinger 1994a)
and, as our group
analysis will show, a deficit of rich clusters compared with the CfA
data.  Defenders of CDM argue that this is not a failing of the theory, but
a consequence of the lack of dissipative baryonic
matter (gas) in the simulation.  Were gas present, it would radiatively
cool and collapse into the dark matter potential wells.   Simulations
including the effects of gas dynamics show that the
cross section of the resulting gas clumps is small enough to significantly
reduce the rate of their mergers (Evrard, Summers, \& Davis 1993).
Katz {et al.} (1992) used dark matter plus gas simulations
to show that very massive dark matter halos can contain several
luminous galaxies, but that smaller halos contain only one galaxy.

We attempt to deal with the overmerging problem in two ways.
The simple approach which enables comparison of our raw simulated catalog
to the CfA data is to ``overmerge'' the real data.  The HG group finding
algorithm
we describe in section 3 finds no simulated groups with more than 13 members
in ten $6^{\circ}$ slices of sky, while the real data, covering only one
$6^{\circ}$
slice, have 3 groups with more than 14 members.  We replace each of these
rich groups in the data with a single galaxy at the group center.
Presumably, a group like this would have merged in the simulation.
We refer to the resulting catalog as the overmerged CfA data.
See Table \ref{galcatalogs} for various properties of the overmerged catalog.

Instead of ``overmerging'' the real data, we can attempt to ``unmerge''
or break up the massive halos in the simulation, using a procedure
like that of Gelb (1992) and Gelb \& Bertschinger (1994b).  The result of this
procedure will be our
breakup catalog.  Properties of this halo catalog are given in Table
\ref{galcatalogs}.

First, we select a maximum $V_{\rm circ}$ for individual halos, then break up
all halos with larger $V_{\rm circ}$.
Assuming a constant cluster mass to light ratio, $M/L$, we convert each
large halo's bound mass to a blue luminosity $L_C$.  The total luminosity
and the number of galaxies brighter than $L$ in a volume $V$ are given by
\begin{equation}
L_{tot}=V \int_0^\infty L \phi(L)dL \,,
\end{equation}
\begin{equation}
N(>L)=V \int_L^\infty \phi(L)dL \,.
\end{equation}
We take $\phi(L)$ here to be the luminosity function of the CfA data,
although in principle $\phi(L)$ could be different for the clusters
and for the field.  Colless (1989) and Schechter (1976)
claim that the luminosity functions of clusters and the field
agree to within their uncertainties.
Eliminating $V$ between these two equations and using the gamma function
$\Gamma(a)=\int_0^\infty t^{a-1} e^{-t} dt$ and the incomplete
gamma function $\Gamma(a,x)=\int_x^\infty t^{a-1} e^{-t} dt$
gives an expression for the number of galaxies brighter than $L$ in
a cluster of total luminosity $L_C$,
\begin{equation}
N(>L,L_C)={L_C \over L^*} {\Gamma(1+\alpha,L/L^*) \over \Gamma(2+\alpha)} \,.
\end{equation}
The halo is then replaced by many luminous objects, with the luminosity
of the $i$th brightest given by $N(>L_i,L_C)=i$.
Positions and velocities for the added
cluster members are selected by randomly choosing a particle
from the halo.

Observations of X-ray clusters reveal the ratio of specific kinetic energy
of the galaxies to that of the gas to be between 0.8 and 1.2 (Sarazin 1988;
Evrard 1990; Lubin \& Bahcall 1993).
We explored scaling the velocities of the added cluster members by
factors in this range, so that the resulting ratio of specific kinetic
energies in galaxies and dark matter particles was between 0.8 and 1.2.
The effects of this scaling should be evident in the ``finger-of-god''
elongation of contours of constant $\xi(r_p,\pi)$.
We found the contours to be insensitive to our scaling.  We conclude
that scaling the velocities of the added cluster members is not necessary,
and we proceed without rescaling.

The choice of $M/L$ used to determine the total luminosity
of the massive halos which are to be broken up also fixes
the luminosities of the added ``cluster'' members.
We preserve the overall form of the luminosity function
by scaling the luminosities of the remaining halos so that they
fill in the gaps left after the added cluster members have been considered.
The net result is an identical luminosity function for the clusters
and the overall catalog.  The mass to light ratio is constant by
construction in our broken up clusters, but varies for other halos
in order to force the correct luminosity function.  For the remainder of this
section $M/L$ will refer to the constant parameter used in the breakup
procedure.

The two parameters we must decide upon are the cluster $M/L$ and the
value for $V_{\rm circ}$ above which we apply our breakup procedure.
The choice of $V_{\rm circ}$ controls the number of broken-up clusters,
while $M/L$ determines the number of added members per halo.
We choose these parameters to make our breakup catalogs statistically
resemble the CfA data as much as possible.  Specifically, we attempt
to match the correlation functions $\xi(s)$ and $\xi(r_p, \pi)$.
While the latter emphasizes the distinction in redshift space clustering
between the radial and transverse directions,
the former function is less noisy
and is useful for characterizing the overall amplitude of the clustering.

Gelb (1992) argues that there are too many halos with $V_{\rm circ} > 350$ km
s$^{-1}$
in the simulation.  We tested breakup $V_{\rm circ}$ values of 300, 350
and 400 km s$^{-1}$, and found no significant difference in the correlation
functions $\xi(s)$ of each.  For all further applications of this
breakup procedure, we adopt a critical $V_{\rm circ}$ of 350 km s$^{-1}$.

Varying
$M/L$ does vary $\xi(s)$ significantly, with lower values of $M/L$
corresponding to higher correlations.  This occurs because a lower $M/L$
means more cluster members are added in a small volume
around the largest halos.  $\xi(s)$ is shown for different values
of $M/L$ in Figure \ref{ic3.2x}.
In order to choose the optimal value for $M/L$ we test the values $250h$,
$500h$ and $1000h$ and compare the resulting
correlation functions and group richness distributions to the observations.
The group richness distribution gives the number of groups as a
function of the number of members in the group.  We have found from these
comparisons that much of the difference between the observations
and the simulation is due to the presence of the Coma cluster in
this particular slice of the CfA data.
If group
richness follows a reasonably smooth distribution function, Coma
is certainly far out on the high end tail, having over 4 times the number
of members of the next richest group.  And, since we are primarily
concerned here with identifying small and medium size groups, we have
no need of simulating an object the size of Coma.  Therefore
it is sufficient that the richness distributions of our simulated catalogs
resemble that of the ``overmerged'' CfA data.

We consider first the
redshift space correlation function $\xi(s)$.  Figure \ref{ic3.2x}
shows $\xi(s)$ for the simulated catalogs and for the CfA data, both
with and without the Coma cluster.
Matching the observed $\xi(s)$ for the full CfA slice data
requires $M/L < 250h$.  This value is smaller than the median $M/L$
found for the nearby groups of HG82, the groups in the first CfA
survey (GH83, Geller 1984), and the groups of Gott \& Turner (1977)
and deVaucouleurs (1975), although RGH89 found a median $M/L = 178h$.
Before breakup, our observed groups (with
a range of mass to light ratios) have a
significantly higher median mass to light ratio (discussed in Paper II).
Using $M/L=250h$ to break up large
halos leads to far too many rich groups, and pie slice diagrams of the
resulting halo distribution show ``fingers of god'' which are far
too prominent to match the observations.
Figure \ref{ic3.2x} reveals that much of the amplitude of $\xi(s)$ for
the complete CfA data is due to the presence of the Coma cluster.
Since we are concerned with finding small groups, it is sufficient
that our simulated data match the clustering properties of the real
data with Coma subtracted.  Using our breakup procedure with
$M/L = 500h$ provides such a match.

The more general two point correlation function $\xi(r_p,\pi)$ separates
the contributions of the radial and the tangential coordinates which
were combined in the separation distance $s$ in $\xi(s)$.  Since
$s^2 \approx (H_0 r_p)^2 + \pi^2$, $\xi(s)$ is roughly equivalent to circular
averages of $\xi(r_p,\pi)$.
Figure \ref{ic3.1x} shows
clearly the extension of the contours in the velocity dimension.  The degree
of this extension in the real data is best matched by both the $M/L=500h$ and
the $M/L=1000h$
simulated catalogs.
Because the group finding algorithms make use of velocity
information and projected separations separately,
matching $\xi(s)$ is not sufficient.  We should strive for the best match
of $\xi(r_p,\pi)$ between our simulated catalogs and the observations.

Our final point of comparison between the breakup catalogs and the observations
is in the group richness distribution.
To make this comparison, groups are identified in both the
simulated and real catalogs, using the HG algorithm
described below in section 3.
Figure \ref{ic3.3x} shows the richness distributions
obtained for the simulations and the real data.
We find that
$M/L=500h$ and $M/L=1000h$ give the best agreement with real data,
while $M/L=250h$ leads to far too many rich groups and clusters.

In summary, simulated breakup catalogs using either $M/L=500h$ and $M/L=1000h$
match the observations reasonably well based on
the multiplicity function and $\xi(r_p,\pi)$ tests, while the $\xi(s)$ test
clearly favors using $M/L=500h$ to break up the largest halos.
The breakup procedure is insensitive to
$V_{\rm circ}$
in the range tested.  Our choice
of parameters for our breakup catalog are therefore $V_{\rm circ}=350$ km
s$^{-1}$
and $M/L=500h$.
By breaking up the artificially large halos in our simulation in this way,
we construct simulated redshift catalogs which match reasonably well
the clustering of the CfA data on scales relevant to small groups.

With the help of G. Tormen, we explored one other option for
``improving'' our simulated catalogs.  A visual inspection of the
slice diagrams reveals that the real data contain larger voids
than does the simulation (cf. de Lapparent, Geller \& Huchra 1986).
This, as well as the low amplitude
of $\xi(s)$ in the simulation, may be due to the absence
of density perturbations on scales larger than the simulation volume.
Tormen \& Bertschinger (1994) have developed an algorithm for
adding long wavelength power on scales which remain linear
to a simulation after it has been evolved gravitationally.

Because the added waves are long, halos which are close together, as
in groups, will all receive approximately the same displacement.
As such, this technique does not disrupt existing groups.  It may,
however, cause halos or whole groups at different initial
locations to move together, creating more or larger groups and
larger voids.  We
generated group catalogs from the wave-added simulation and
found no significant difference in the groups.  Using the same
halo as an observation point, the wave-added and non-wave-added
slice diagrams and group catalogs appeared more like different
regions of space in the same universe than like regions with fundamentally
different clustering properties corresponding to distinct universes.
We take this as an indication that groups are not particularly
sensitive to the size of nearby voids.
For this reason we chose not to include separate analyses of
groups in the wave-added simulation.

\section{Group Finding Algorithms}
We consider two grouping algorithms.  Each is
a friends of friends type algorithm with variable linking lengths.
They differ only in the scaling of the linking parameters.
For each algorithm, we first describe its implementation, then
apply it to
our simulated galaxy catalogs,
making use of only the
observationally available right ascension, declination, and redshift.
We test the resulting group catalogs for
the accuracy of group membership assignments.

Both group finding algorithms studied here operate by considering
whether or not each pair of galaxies is linked, according to some
specified criteria.  Because of the uniqueness of the radial
coordinate in redshift surveys, both algorithms adopt two linking
criteria: one for the redshift separation between two galaxies
or halos, and one for the transverse (projected) separation.
For each, a linking length which varies with redshift is
used for reasons discussed below.  Galaxies or halos are considered linked if
their separation
in both the transverse and radial dimensions is less than the corresponding
linking length.  Groups are identified as collections of mutually
linked galaxies or halos.

Because we will be referring to group catalogs calculated in different ways,
it will be convenient for us to define some abbreviations.  The basic
application of the group finding algorithms we test here
is to a galaxy or halo catalog which is complete to some apparent magnitude
limit and which gives galaxy or halo redshifts, but not true distances.
We refer to group catalogs constructed from these halo catalogs
by the letters Vm, with
V referring to the use of velocity (redshift) as the radial coordinate and m
indicating
that the groups were found by searching an apparent magnitude limited
catalog.  If, as in the simulated data, we know true distances,
we can apply the grouping algorithms using true distance instead of
redshift distance as the radial coordinate.  The resulting group
catalog we label Rm, where R instead of V means we have used true distance
instead of velocity in the group finding algorithm.  Comparing
these two types of group catalogs tells us how peculiar velocities
affect group identification.  Finally, we also construct group catalogs
from an absolute magnitude limited sample of halos.  These are our
RM catalogs.  The capital M refers to an absolute magnitude limited
halo catalog, while the lower case m refers to an apparent magnitude limit.
These labels refer either to a halo catalog or, when describing
a group catalog, to the halo catalog from which the groups were
identified.
We probe the effects of peculiar velocities on group identification by
comparing Vm to Rm catalogs.  Any effects or biases in group properties
due to the absence of faint groups in the Rm catalog can be studied
by comparing it to the RM catalog.  By using more complete information
as we go from Vm to Rm to RM catalogs, we separate the effects
of peculiar velocities and the flux limit.
Table \ref{cattypes}, describing the different catalog types,
is provided for reference.

Our Rm catalogs are constructed according to the group finding algorithms
described below, but using true distances instead of redshift distances.
Because groups and
clusters are not artificially extended along the line of sight in real space,
the radial
linking length and transverse linking length are identical (but
distance-dependent).
Our other real space (RM) catalogs are constructed using a fixed linking
length in an absolute magnitude limited galaxy sample.
This technique
finds the faint groups missed in the apparent magnitude limited catalog.
It tests whether the variable linking length used to construct groups
from the apparent magnitude limited catalogs is too large.

Each of our catalogs covers a portion of sky identical in geometry to the
first slice of the CfA survey extension: a 9 hour range
in right ascension and declination between $26.5^\circ$ and
$32.5^\circ$.  For each of our simulated group
catalog types (Vm, Rm and RM), ten catalogs were generated, corresponding to
ten
different ``observers.''  Observer galaxies were chosen to be the ten
halos just fainter than $L^*$ with peculiar velocities between about 350 and
650 km s$^{-1}$.
Of these ten observers, two were within about $10h^{-1}$ Mpc of an extremely
massive
halo of the type which exists due to the overmerging problem.  Other than
these two, none of the observer halos is in any kind of extreme environment.
Unless otherwise noted, all statistics given for the group catalogs
will refer to averages for the ten group catalogs generated by
the ten observers.  All in all, we have 10 realizations (observers)
of 2 versions, raw and breakup, for a total of 20 simulated halo catalogs,
each of which can be used to construct a Vm, Rm and RM group catalog.
Because they more closely match the observed halo clustering and
group multiplicity functions, we prefer our breakup catalogs to the raw ones.

\subsection{The Huchra \& Geller (HG) Algorithm}
HG82 (also GH83 and RGH89) sought to
identify groups as number density enhancements
in redshift space.  They employed a friends-of-friends algorithm with two
variable linking lengths, one for projected separation and one for
the redshift dimension.  Specifically, they declared a pair of galaxies
to be linked if their projected separation and velocity difference
are less than or equal to certain critical values,
\begin{equation}
D_{12}=2\sin (\theta /2)V/H_0 \leq D_L(V) \,, \nonumber\\
\end{equation}
\begin{equation}
V_{12}=|V_1-V_2| \leq V_L(V) \,,
\end{equation}
where $V_1$ and $V_2$ are the line of sight velocities of the
two galaxies, $V=(V_1+V_2)/2$, and $\theta$ is the pair's angular separation.
All pairs of galaxies were searched for linkage, and each disjoint
set of linked galaxies was declared a group.

HG82 scale the transverse and radial linking lengths
to compensate for the decline of
the selection function with distance:
\begin{eqnarray}
D_L&=&D_0{\left[\int_{-\infty}^{M_{V}} \phi(M)dM \bigg/
\int_{-\infty}^{M_{\rm lim}} \phi(M)dM\right]}^{-1/3} \,, \nonumber\\
V_L&=&V_0\left[\int_{-\infty}^{M_{V}} \phi(M)dM \bigg/
\int_{-\infty}^{M_{\rm lim}} \phi(M)dM\right]^{-1/3} \,,
\label{dl}
\end{eqnarray}
where $M_{V}=m_{\rm lim}-25-5\log(V/H_0)$ is the absolute magnitude
of the brightest galaxy visible at a distance $V/H_0$.  Similarly,
$M_{\rm lim}=m_{\rm lim}-25-5\log(V_F/H_0)$ is the absolute magnitude of
the brightest visible galaxy at a fiducial distance $V_F/H_0$.
$D_0$ and $V_0$ are the linking cutoffs at $V_F$, and $\phi(M)$ is
the galaxy absolute magnitude (luminosity) function.
Assuming groups are spherical, this corresponds to selecting for
a minimum excess number density
\begin{equation}
\bigg({\delta \rho \over \rho}\bigg)_{\rm crit}={3 \over 4\pi D_0^3}
\left[\int_{-\infty}^{M_{\rm lim}} \phi(M)dM\right]^{-1} -1.
\label{overdencrit}
\end{equation}
We use the abbreviation HG to refer to this algorithm and to the
groups identified in this manner.

We generate our Vm group catalog by applying the HG algorithm
to our simulated catalogs
using the search parameters of RGH89.  These are
\begin{equation}
D_0=0.27h^{-1} {\rm~Mpc}, \qquad V_0=350 {\rm~km s^1}, \qquad V_F=1000 {\rm~km
s^1} \,,
\label{lparam}
\end{equation}
which give a critical overdensity $(\delta \rho / \rho)_{\rm crit}=80$.
In order to handle the few blueshifted galaxies, velocities less
than 300 km s$^{-1}$ are set to 300 km s$^{-1}$, and although individual
galaxies
can have velocities up to 15000 km s$^{-1}$, groups with mean velocities
greater than 12000 km s$^{-1}$ are excluded.  HG82 and GH83 justify this last
step by arguing that at these large velocities, uncertainties in the
bright end of the luminosity function translate into large uncertainties
in the scaling of the linking lengths.

We also locate groups using true position information (Rm catalog),
by modifying the linking
condition in the HG algorithm. Here two galaxies are
considered linked if their true separation is less than $D_L$, as
defined in equations (\ref{dl}), in both the projected separation and the
radial
separation.  We accomplish this by setting $V_L$ equal to $H_0 D_L$.
$V_L$ need not be large in this real space linking, since
groups do not appear extended along the line of sight in true distance.

The first test of redshift space effects on the HG groups is to see how well
the HG algorithm in redshift space
reproduces the group catalog determined using true distances.  A visual
check can be made comparing the simulated Vm and Rm catalogs.
Figure \ref{iia1.6x} shows four $6^{\circ}$ thick pie
slices of the simulated breakup sky, plotted using either velocity or true
distance
as the radial coordinate.  For each space (real or redshift), breakup groups
identified using either velocities or true distances are shown as
sets of crosses linked to their geometric center.
The equivalent information is shown in Figure \ref{iia1.1x} for the
raw groups.
It is apparent from these figures that many more groups are found in redshift
space than in real space.  This is because the radial linking length
is larger in the redshift linking case by a constant factor of
$V_0/D_0 \approx 13$.
The prominent redshift space ``fingers'' formed by groups which are quite
compact in real
space argue for the necessity of a large linking length in the radial
dimension.  Each finger in Figure \ref{iia1.6x}(b) corresponds to
one compact group in Figure \ref{iia1.6x}(a).
The necessity of this generous velocity linking length can be seen even
when the breakup procedure is not employed, by
comparing the 6 member group at 3 hours right ascension and
$cz=11000$ km s$^{-1}$ in Figures \ref{iia1.1x}(a) and
\ref{iia1.1x}(b).
Although this group is quite extended in redshift space, it is actually
very compact.  In order to locate such groups in redshift space, a large
velocity linking distance is required.  The price we pay for including
high velocity dispersion groups is the inclusion of many non-physical
associations caused by projection.  It is obvious by inspection of Figures
\ref{iia1.6x}(c) and
\ref{iia1.1x}(c) that many of the groups identified in redshift space
are these non-physical projections.

We quantify the accuracy of group memberships by
counting the number of members of a redshift identified (Vm) group
which belong to a common real space group (Rm or RM).  For this purpose we
include binary galaxies in our real space groups.
We first determine which members of each Vm group belong to the same Rm or RM
group.  We calculate a largest group fraction, or LGF, by dividing the
number of halos in the largest such subgroup by the total number of members
in the Vm group.
Figure \ref{lgf.hgVmRm} shows, as a function of group richness,
the fraction of groups of a given richness $N$ with LGFs of unity and in each
quartile below.
The total number of groups with $N$ members is also given.
For example, there are 26 groups with 7 members.  Of these, one half have LGFs
of 100\%, 85\% have LGFs of 75\% or more, 92\% have LGFs of at least one half,
and all of the $N=7$ groups have LGFs greater than 25\%.

The Vm to Rm comparison in Fig. \ref{lgf.hgVmRm} tells us how much more
accurate
the grouping algorithm could be if we had accurate distance indicators.
Only 107 of the 222 triplets found
in redshift space actually belong together in real space.  Fourteen
of those 107 groups were identified with $N \geq 5$ in real space; the rest
were
either triplets or quartets in real space.
However, although only about one half of the $N=3$ groups correspond to
$N \geq 3$ real space (Rm) groups, another 35\% of the redshift
space (Vm) triplets contain a pair of halos which are linked in
real space.  Thus almost one half of triplets are accurately identified
and another third are actually binaries with one interloper.
The remainder are either complete misidentifications or correspond
to picking one member from a real space group and adding two interlopers.

This result is roughly consistent
with the claim by RGH89 that one third or more of the $N=3$ groups
in their sample
are spurious.  They do not, however, distinguish between the
binary galaxies with one interloper and completely erroneous
identifications.  RGH89 also claim that almost all of their $N=5$ and
richer groups are real.  Our simulations do not fully support this conclusion.
Although Figure \ref{lgf.hgVmRm} shows that almost all of our $N=5$ groups
correspond to real space (Rm) groups of three or more, only about half
of them
contain no interlopers.
This is not a concern when studying the
spatial distribution of all $N \geq 3$ groups, since finding an $N=5$ group
almost certainly means a true group of three or more members occupies that
position.  It may be a concern, however, when calculating internal
group properties such as virial mass estimates, since the presence
of interlopers in half of the $N=5$ groups is expected to bias the results.
For groups richer than $N=5$, again, almost all correspond to a real space
triplet or greater.  In general, the more members the Vm group contains,
the higher the accuracy of the group membership.

Since RGH89 selected $N \geq 5$ groups for their
determination of median group properties, our findings call
their results into question.
Our tests confirm their result that $N=5$ groups are very rare in an
unclustered galaxy distribution.  However, that is not reason enough
to trust that $N=5$ groups in a clustered distribution are
accurately identified.
The likely presence
of interlopers in the RGH89 groups may bias their derived group properties.
We investigate this possibility in Paper II.

The most accurate group identification scheme hypothetically
possible requires that we know true distances for an incredibly
deep (in magnitude) sample which could then be absolute magnitude
limited.
We calculated LGFs based on comparing Vm to RM groups and found their
distribution to be very similar to that of the Vm to Rm comparison
shown in Figure \ref{lgf.hgVmRm}.  This result is promising,
since it means that the accuracy of the algorithms can be studied
even though faint groups and members are absent at large distances.

Since our specific aim in locating groups was the identification
of regions with an excess number density of galaxies, another accuracy
check is to calculate the overdensities for the Vm groups.  These
are calculated according to the formula
\begin{equation}
{\delta \rho \over \rho}={N \over V} \bigg/
\int_{-\infty}^{M_{V}} \phi(M)dM - 1 \,,
\label{overden}
\end{equation}
where $N$ is the number of member halos and $V$ is the volume
enclosing the group, estimated roughly
as the volume of an elliptical cylinder encompassing the
members,
\begin{equation}
V=\pi \bigg({V_{max}+V_{min} \over 2H_0} \bigg)^2
\bigg({\alpha_{max} - \alpha_{min} \over 2}\bigg)
\bigg({\delta_{max} - \delta_{min} \over 2}\bigg)
\bigg({V_{max} - V_{min} \over H_0}\bigg) \,.
\end{equation}
Note that this is a measure of the overdensity in redshift space, and
therefore does not reflect the true compactness of a group.  In fact,
because it is proportional to the maximum velocity difference in the group,
truly dense groups with high velocity dispersions will have artificially low
overdensity values.

Figure \ref{iia1.4x} shows this overdensity statistic plotted against LGF
for our breakup groups.
There is a generally increasing trend in LGF with overdensity, indicating
that the most compact groups are the most accurately identified.
Figure \ref{iia1.5x} shows LGF as a function of group redshift.
A clear trend of decreasing LGF with redshift is evident.
This occurs because the radial linking length is much larger for the
Vm groups than for the Rm groups.  As a result, distant Rm groups
with similar angular positions but different redshifts are more likely
to be linked into one Vm group, with a low LGF.

Although our raw halo catalogs resemble the real data less strongly
than do our breakup catalogs, we have nevertheless performed our accuracy
checks on raw groups also.
Many of the groups in the breakup catalogs correspond to single
massive halos in the raw catalog.  Those that are isolated
do not appear as members of raw groups.  Also, because of the
conservation of luminosity condition of the breakup procedure,
each small halo is more luminous in the raw catalog than in the breakup
catalog.  In fact, many of the faint raw halos fall below the
magnitude limit after breakup is performed.  The net result of these
differences is that groups in the raw catalogs tend to be less rich
and less dense than breakup groups.
This can be seen initially
in the number of Rm groups found in the raw catalogs (see Fig. \ref{iia1.1x}).
There are almost 3
times as many Vm as Rm groups in the raw catalogs.  In the breakup case,
the difference is just under 20\%.

The LGFs for the raw catalogs are lower than in the breakup case,
as is clear from
Figure \ref{lgf.hgVmRm}.
This is due to the fact that the accuracy of the breakup groups
is large partially as a result of the breakup procedure; broken
up halos, being compact in real space, appear as very accurately
identified groups.  Raw groups do not have this ``advantage.''

Like the groups in the breakup catalog, the raw groups exhibit
a trend of increasing LGF with overdensity.
The amplitude of the
relation is higher for the breakup case, with LGFs approaching
unity at the highest overdensities.
As a function of redshift, however, raw and breakup groups behave
differently.
While the breakup groups tend to be less accurately
identified at larger redshifts, Figure \ref{iia1.5.raw} indicates that
the LGFs of raw groups actually rise
with redshift for the Vm to Rm comparison and are almost flat for the
Vm to RM comparison.
The trend here is opposite the trend for the breakup groups because for the raw
groups, the triplets are actually more likely to have a high LGF than
are the richer groups, since most of the richer raw Vm groups are actually
poor Rm groups which have been linked.  Since triplets dominate at
large redshifts due to the apparent magnitude limit,
LGF increases with redshift.
We expect that because the breakup halo catalogs more closely
match the clustering properties of the real data, the trends evident
in the breakup case are more likely to represent reality.

\subsection{Maximizing Group Accuracy}
Now that we have developed the tools for quantifying the accuracy
of the group finding algorithms, we can explore the $(D_0, V_0)$
parameter space to try and maximize group accuracy.  We study
this space on a grid of values with $D_0$ equal to $0.4233h^{-1}$,
$0.3387h^{-1}$, $0.2700h^{-1}$, and $0.2147h^{-1}~{\rm Mpc}$,
corresponding to critical overdensities
(eq. [\ref{overdencrit}]) of 20, 40, 80 and 160, respectively, and
with $V_0$ equal to 150, 250, 350 and 550 km s$^{-1}$.
At each of the parameter space points, we
constructed raw group catalogs (Vm and Rm) for our ten observers
and calculated LGFs.
We constructed RM catalogs based only on our standard parameters,
$(D_0, V_0)=(0.27h^{-1}~{\rm Mpc}, 550~{\rm km s^1})$, and computed LGFs
also by comparing Vm groups at each parameter space point with these
RM groups.

Table \ref{accuracy} shows the grid
of parameter space values tested, with accuracy statistics at each
grid point.
The top and bottom number at each grid point are, respectively,
the mean LGF from the Vm to Rm comparison and the mean LGF from the Vm to RM
comparison for the breakup simulated groups.
LGFs are apparently sensitive to both $D_0$ and $V_0$, with the mean
LGFs of our breakup groups increasing as the linking parameters are
made more restrictive.  Because our breakup groups tend to be compact in
real space, generous linking lengths allow for the inclusion of more
interlopers than true members.
As a result, our maximum mean LGF occurs at a corner in the parameter
space we explore, at $D_0=0.2147h^{-1}~{\rm Mpc}$ and $V_0=150$ km s$^{-1}$.
Repeating this parameter space search using our raw catalogs reveals the
same trend of LGF with $V_0$, but only an insignificant dependence
on $D_0$.  For looser groups then, the accuracy of the HG algorithm
is sensitive only to the radial linking parameter $V_0$.

This result differs with the claim made by RGH89 that their choice
of linking parameters, our standard parameters, minimizes
the number of interlopers in the group catalog.  We find that the
number of interlopers can be reduced by decreasing the velocity linking length
in the grouping algorithm,
at the price of missing some true group members.

Our results indicate that the optimal set of parameters for reducing
the number of interlopers in HG groups are far from optimal for determining
group velocity dispersions.
Instead, the ideal parameters depend on
the purpose for which groups are being identified.  For example,
if one wishes to study the spatial distribution of groups, it is important
that the choice of $V_L$ is not so restrictive as to cause high velocity
dispersion groups to be missed.
In that case, the parameters used by RGH89, our standard parameters,
are a good choice.  If one wanted to study members of groups
for signs of mergers or close encounters, then a smaller $V_0$,
resulting in a more certain identification of group members, is
superior.  In Paper II we consider the effects of the $V_0$ parameter
on group velocity dispersions and mass estimates.

\subsection{The Nolthenius \& White (NW) Algorithm}
NW87 identified groups in the first CfA redshift
survey, which covered 2.66 steradians to an apparent magnitude limit
$m_{B(0)} = 14.5$.  Their algorithm differs from the HG algorithm
only in the scaling of the two linking lengths, $D_L$ and $V_L$.

NW87 argue that the velocity scaling of the HG algorithm increases too rapidly
at large redshift.  The velocity scalings are compared in Figure \ref{iib1.1x}.
For the Schechter function parameters
given in equation (\ref{schech}), $V_L=3V_0$ at a distance of 8300 km s$^{-1}$.
They reason that groups at larger redshifts will be brighter (since
fainter groups fall below the survey limit) and hence will have
a higher typical velocity dispersion.
Since they believed that $V_L$ should be large enough to not bias the group
velocity
dispersions and not appreciably larger, NW87 chose to scale their $V_L$
with distance in the same way the typical group velocity dispersion
varies with distance.  They located groups in real space in their
simulation using a
three dimensional linking parameter scaled using the HG scaling,
and found that group velocity dispersions increased linearly with
distance.  Specifically, they scaled $V_L$ as
\begin{equation}
V_L=V_0+0.03(V-5000 {\rm km s^1}) \,.
\label{nwvl}
\end{equation}
Based on scaling arguments we will not repeat here, NW87 scale $D_L$ as
\begin{equation}
D_L=D_0 \left[ {\int_{-\infty}^{M_{V}} \phi(M)dM \over
\int_{-\infty}^{M_{\rm lim}} \phi(M)dM} \right]^{-1/2}
\left[{V_F \over V} \right]^{1/3} \,,
\end{equation}
where $M_{V}$ and $M_{\rm lim}$ are defined as for equation (\ref{dl}).
We refer to the group finding algorithm employing these scalings as the
NW algorithm, and to the resulting groups as NW groups.
As can be seen in Figure \ref{iib1.1x}, the HG and NW scalings of $D_L$
are quite similar.

Figure \ref{iib1.3x} shows slices of sky, analogous to Figure \ref{iia1.1x},
depicting groups identified from one of our breakup catalogs in
redshift space (Vm) and in real space (Rm)
and plotted in both redshift and real space.  The scaling of the
velocity linking length seems to be too restrictive here, since some
groups identified in real space are not found by the redshift space
algorithm.  Note the two Rm groups at about 3 hours and 11,000 km s$^{-1}$
which
do not appear as Vm groups.

Displayed in Figure \ref{lgf.nwVmRm} are
the largest grouped fractions (LGFs) for our breakup NW groups
When grouped
by number of members as in Figure \ref{lgf.hgVmRm}
the LGFs for the breakup NW groups are
larger than those of the breakup HG groups.  There are more NW groups
with LGFs of unity than for the HG groups.
Because it uses a smaller velocity linking length, the NW algorithm
finds fewer interlopers than the HG algorithm, resulting
in the higher LGFs.

Plots of mean LGF vs. overdensity and redshift, not shown here, display
similar behavior for the NW groups as for the HG groups.
For the linking parameters tested here, the NW algorithm does not
identify the low density groups found by the HG algorithm.
As a function of overdensity, LGFs for the raw NW groups show
behavior similar to that of the HG groups in the same overdensity range.

As in the case of the HG algorithm, breaking up the large halos
increases the accuracy of the groups.  LGFs
are similar for the NW and HG algorithms when applied to the raw
catalog.
In fact, the only significant difference between the NW and HG algorithms
applied to the raw catalogs is that the NW groups have a higher
overall amplitude of their LGFs.  This is consistent with our
result for the breakup catalogs, where NW groups had higher LGFs.

As with the HG algorithm, the optimal set of linking parameters to use
depends on the purpose for which groups are being identified.
Optimizing the NW algorithm requires the exploration of an additional
dimension in parameter space, the slope in the NW velocity scaling in
equation (\ref{nwvl}).  Following NW87, we calculated the average
group velocity dispersion as a function of distance in our Rm groups.
The slope used by NW87, 30 km s$^{-1}$ in dispersion for every 1000 km s$^{-1}$
in distance,
holds reasonably well for our groups also.  However, this value is an
unweighted average with a large variance, indicating that many of these
groups have significantly higher velocity dispersions.  We have already
seen that the NW algorithm (with NW87's parameters) misses real groups
with large velocity dispersions.  If that is unacceptable to a user
of the NW algorithm, he or she may increase either the normalization
($V_0$) or the slope in the NW velocity scaling.

It is worth noting that in recent work on
groups in the original CfA survey, Nolthenius (1993) chose to use
the HG velocity link scaling, arguing
that the NW linear scaling resulted in a negative correlation between
group $M/L$ and redshift, while the HG scaling did not.
Since the NW and HG $D_L$ scalings are similar (Fig. \ref{iib1.1x}),
our discussion in section 3.2 on optimizing the linking parameters for the HG
algorithm should apply to this combined algorithm also.

\section{Summary and Conclusions}
We have generated simulated galaxy redshift surveys from
numerical experiments and used them to test group
finding algorithms in redshift space.  Construction
of the simulated catalogs required the development
of our latest version of the DENMAX halo identification
algorithm, as well as extensive testing to determine
the best method for illuminating the halos in the simulation
and dealing with the overmerging problem.  We also tested
whether the group finding algorithms are sensitive to
the absence of power on scales larger than the fundamental
simulation volume
and concluded that they are not.

The HG and NW algorithms for group identification both attempt to
locate groups of galaxies whose number density in redshift space
is above some threshold.  Because they lack true distance
information, neither can be expected to be perfect;
each must strike a balance between identifying false groups
and missing real ones.  The HG algorithm leans toward the former.
Almost every group identified in a magnitude limited catalog with
true distances is also found by
the HG algorithm in redshift space, often with added members.  The
HG procedure also finds many completely spurious small groups.  The NW
algorithm, on the other hand, identifies fewer false groups but also
misses some real ones.  This difference may help future
workers decide which algorithm best suits their purpose.  We note
that the differences between the HG and NW algorithms in terms of
membership accuracy are stronger for our raw catalogs, which are less
similar to the real data than are our breakup catalogs.
Searches for evidence of mergers or interacting galaxies
in groups would benefit from
high accuracy in group
identification, a feature of the NW algorithm, while calculating
the spatial distribution of groups requires that few be missed, as
with the HG algorithm.

Our attempts to optimize
or fine tune the parameters of the group finding algorithms
reveal that the optimal parameters, like the choice of HG
or NW algorithms, are dependent on the purpose for which
groups are being located.  Restrictive velocity linking
lengths in either the HG or NW algorithms cause high
velocity dispersion group members to be missed but result in
fewer interlopers.
Once the decision is made as to which algorithm is more
appropriate to a particular purpose, one must also select
linking parameters which best suit that purpose.
Our numerical simulations provide a good way to make these choices.

The overmerging problem present in our simulation and others like it
causes what presumably should be compact groups and clusters to
collapse into single massive clumps.
The accuracy of both HG and NW algorithms depends sensitively
on whether or not we break up the overly massive halos in our
simulation.  Groups resulting from our breakup procedure
are identified significantly more accurately than the raw groups.
Breakup also results in many more rich
groups and fewer poor groups.  We find that the accuracy of our breakup
and raw groups differ in their dependences on redshift, group richness
and the transverse linking length employed in the group finding
algorithms.  Breakup group accuracy correlates with richness and anticorrelates
with $D_0$ and redshift; raw groups display the opposite behavior.
We take these differences to mean that the overmerging problem
must be dealt with carefully if dissipationless simulations are to be
used to study galaxy groups.

Our results roughly confirm the claim by RGH89 that approximately
one third of $N=3$ groups are false, but indicate that their claim
that $N \geq 5$ groups are accurate must be qualified.  We find that
groups richer then $N=5$ almost all contain a real group of three or more
members.  However, many of these rich groups are not completely
accurate, having been contaminated by interlopers.
In Paper II we show that although this affects the accuracy of individual
group masses, the distribution of group masses is less sensitive.

Future studies of galaxy groups will require that care be taken
to balance completeness considerations with the problem of interlopers
when identifying groups in redshift catalogs.

\acknowledgments
Supercomputer time was provided by the Cornell National Supercomputer
Facility and the National Center for Supercomputing Applications.
Support was provided by NSF grants AST90-01762 and ASC93-18185
and by an NSF graduate
fellowship.  The author is grateful for the guidance of Ed Bertschinger
and the previous work of Bertschinger and James Gelb.

\clearpage

%Tables here.
\begin{table}
\centering
%\centerline{Table 1: Galaxy Catalog Properties}
\caption{Galaxy Catalog Properties}
\vspace{12 pt}
{\small
\begin{tabular}{lccccc}
\hline
\hline
& $N_{\rm halos}$& $\langle cz \rangle_{\rm med}$& $s_0$& $-\gamma$&
$\sigma_{v, {\rm pec}}$\\
Catalog& & (km s$^{-1}$)& $(h^{-1}~{\rm Mpc})$& & (km s$^{-1}$)\\
\hline
CfA& 1094& 7405& $8.68 \pm 0.05$& $1.09 \pm 0.01$& ...\\
Overmerged CfA& 913& 7544& $4.10 \pm 0.03$& $1.31 \pm 0.01$& ...\\
Raw& $917 \pm 69$& $6452 \pm 789$& $2.72 \pm 0.03$& $1.25 \pm 0.02$& $437 \pm
47$\\
Breakup& $890 \pm 116$& $6705 \pm 904$& $3.90 \pm 0.03$& $1.30 \pm 0.02$& $482
\pm 47$\\
\hline
\end{tabular}
}
%$^a$~Redshifts and $\sigma_{v,pec}$ are given in km s$^{-1}$
%
%$^b$~From a power law fit to $\xi(s)$ over the range $1h^{-1} \leq s \leq
%%10h^{-1}$ Mpc
%
%$^c$~Quantities are mean $\pm$ one standard deviation for our ten simulated
%%catalogs
%\caption{Galaxy catalog properties:  Redshifts and peculiar velocity
%dispersions $\sigma_{v,pec}$ are given in km s$^{-1}$.  Correlation function
%fitting parameters $s_0$ and $\gamma$ are from a power law fit to
%$\xi(s)$ over the range $1h^{-1} \leq s \leq 10h^{-1}$ Mpc.
%For the raw and breakup halo catalogs, quantities quoted are mean
%$\pm$ one standard deviation for our ten simulated catalogs.}
\vspace{12 pt}
\label{galcatalogs}
\end{table}

\begin{table}
\centering
%\centerline{Table 2: Group Catalog Types}
\caption{Group Catalog Types}
\vspace{12 pt}
\begin{tabular}{lll}
\hline
\hline
Catalog Type & Radial Coordinate & Halo catalog limit \\
\hline
Vm & Velocity & Apparent magnitude \\
Rm & True distance & Apparent magnitude \\
RM & True distance & Absolute magnitude \\
\hline
\end{tabular}
%\end{center}
%\caption{Group catalog types.}
\label{cattypes}
\end{table}

\begin{table}
\centering
%\centerline{Table 3: Group Accuracy as a Function of $D_0$ and $V_0$}
\caption{Group Accuracy as a Function of $D_0$ and $V_0$}
\vspace{12 pt}
{\small
\begin{tabular}{lcccc}
\hline
\hline
$V_0$ & \multicolumn{4}{c}{$D_0$ ($h^{-1}$ Mpc)} \\
%\cline{2-5}
(km s$^{-1}$) & 0.4233 & 0.3387 & 0.2700 & 0.2147\\
\hline
150&0.83  &  0.84  &  0.86  &  0.88 \\
   &0.66  &  0.71  &  0.77  &  0.83 \\
\hline
250&0.79  &  0.80  &  0.84  &  0.86 \\
   &0.61  &  0.67  &  0.74  &  0.81 \\
\hline
350&0.75  &  0.77  &  0.81  &  0.85 \\
   &0.58  &  0.65  &  0.72  &  0.79 \\
\hline
550&0.69  &  0.73  &  0.78  &  0.81 \\
   &0.55  &  0.62  &  0.69  &  0.76 \\
\hline
\end{tabular}
}
%\end{center}
%\caption{Group accuracy as a function of $D_0$ and $V_0$: The top and bottom
%%number at each grid point are, respectively,
%the mean LGF from the Vm to Rm comparison and the mean LGF from the Vm to RM
%comparison for the breakup simulated groups.  Values of $D_0$ are in $h^{-1}$
%%Mpc and values
%of $V_0$ are in km s$^{-1}$.  The mean is computed over all groups in all
%ten catalogs.}
\label{accuracy}
\end{table}

\clearpage

\begin{figure}
\centerline{Figure Captions}
\caption{Redshift space correlation functions $\xi(s)$ for the
CfA data and averages of $\xi(s)$ for 10 simulated catalogs.
Dotted lines represent CfA data; solid lines are for the simulated data.
Shown are the CfA data with and without Coma and without the 3 largest
groups (all with 25 or more members).  Note the large effect of Coma on both
the amplitude and slope of $\xi(s)$.  For the simulated data
we plot $\xi(s)$ for the raw catalog and for breakup catalogs with
different values of $M/L$.}
\label{ic3.2x}
\end{figure}

\begin{figure}
\caption{Six panels show contours of constant $\xi(r_p,\pi)$.
Solid contours trace the values 1, 2, 4, and 8, and dotted contours
represent values of 0.5 and 0.25.  Panels are: CfA data (a),
CfA with Coma members removed (b), Raw simulation (c); and breakup
catalogs with $M/L=250h$ (d), $M/L=500h$ (e), and $M/L=1000h$ (f).
For the simulations we plot the average $\xi(r_p,\pi)$ for 10 halo
catalogs.}
\label{ic3.1x}
\end{figure}

\begin{figure}
\caption{Number of groups $N_{\rm gr}$ with $N_{\rm mem}$ members for our
raw catalog
and for 3 breakup catalogs.  These groups were found by application
of the Huchra \& Geller algorithm described in section 3.
Error bars give the standard deviation over 10
catalogs.  Stars represent groups in the CfA data.}
\label{ic3.3x}
\end{figure}

\begin{figure}
\caption{Four panels show $6^{\circ}$ thick simulated slices of
``breakup'' sky.
All visible halos in the slice are shown as circles.
Panels (a) and (c) are plotted with true distance as radial coordinate;
panels (b) and (d) use redshift distance.
Breakup HG groups are shown as crosses connected to their geometric center.
Panels (a) and (b) show groups identified in real space (Rm);
panels (c) and (d) show redshift space (Vm) groups.  Radial distance
extends to 15000 km s$^{-1}$.}
\label{iia1.6x}
\end{figure}

\begin{figure}
\caption{Four panels show $6^{\circ}$ thick simulated slices of ``raw''
sky.  All visible halos in the slice are shown as circles.
Panels (a) and (c) are plotted with true distance as radial coordinate;
panels (b) and (d) use redshift distance.
Raw HG groups are shown as crosses connected to their geometric center.
Panels (a) and (b) show groups identified in real space (Rm);
panels (c) and (d) show redshift space (Vm) groups.  Radial distance
extends to 15000 km s$^{-1}$.}
\label{iia1.1x}
\end{figure}

\begin{figure}
\caption{Distribution of largest grouped fraction (LGF) as a function of the
 number of members $N$ in
HG Vm groups.  Left (right) panel shows LGFs of breakup (raw) groups,
based on comparison of Vm groups to Rm groups.
Cross-hatched regions give the percentage of groups with LGFs of unity,
single narrow-hatched regions correspond to groups with LGFs between 75\% and
100\%, single wide-hatched regions represent groups with LGFs
between 50\% and 75\%, and
no hatching represents groups with LGFs between 25\% and 50\%.
The number at the top of each bar is the total number of HG groups with
$N$ members in our ten breakup catalogs.  The $N=10+$ ($N=6+$) bar includes
all groups with ten (six) or more members.}
\label{lgf.hgVmRm}
\end{figure}

\begin{figure}
\caption{Smoothed curve of LGF vs. the logarithm
of group overdensity for the breakup HG groups.
LGF is computed by comparing groups selected from a magnitude
limited redshift survey simulation (Vm) to real-space selected groups
in apparent (Rm) or absolute (RM) magnitude limited samples.
Groups were sorted by overdensity,
then a moving average over 41 groups was performed.
Each point on the ordinate is the mean of the LGFs for the 41 groups
whose median overdensity is plotted on the abscissa.}
\label{iia1.4x}
\end{figure}

\begin{figure}
\caption{Smoothed curve of LGF vs. group redshift
for the breakup HG groups.
The two curves have the same meaning as in Fig. \protect\ref{iia1.4x}.}
\label{iia1.5x}
\end{figure}

\begin{figure}
\caption{Smoothed curve of LGF vs. group redshift
for the raw HG groups.
The two curves have the same meaning as in Fig. \protect\ref{iia1.4x}.}
\label{iia1.5.raw}
\end{figure}

\begin{figure}
\caption{Relative scalings of different linking lengths.
Solid line is the HG scaling appropriate for the luminosity function
used here.  The dashed line is NW's projected
separation scaling, for the same luminosity function.  The dotted
line is NW's linear scaling for the velocity linking length.
All curves are normalized to coincide at 1000 km s$^{-1}$.}
\label{iib1.1x}
\end{figure}

\begin{figure}
\caption{Four panels show $6^{\circ}$ thick slices of sky.
All visible halos in the slice are shown as circles.
Panels (a) and (c) are plotted with true distance as radial coordinate;
panels (b) and (d) use redshift distance.
Breakup NW groups are shown as crosses connected to their geometric center.
Panels (a) and (b) show groups identified in real space (Rm);
panels (c) and (d) show redshift space (Vm) groups.  Radial distance
extends to 15000 km s$^{-1}$.}
\label{iib1.3x}
\end{figure}

\begin{figure}
\caption{Distribution of LGF as a function of the number of members $N$ in
breakup NW Vm groups,
analogous to the left panel of Fig. \protect\ref{lgf.hgVmRm} for breakup HG
groups.}
\label{lgf.nwVmRm}
\end{figure}

\clearpage

\end{document}